\documentclass[UTF8,a4paper]{article}
\usepackage[UTF8, fontset=windows]{ctex}
\usepackage[sort&compress,numbers]{natbib}
\usepackage{booktabs}
\usepackage{geometry}
\usepackage{titlesec}
\usepackage{url}
\usepackage[font={small,bf}]{caption}
\usepackage[yyyymmdd]{datetime}
\usepackage{graphicx}
\usepackage{multirow}
\usepackage{float}
\usepackage{lmodern}
\usepackage{hyperref}
\titleformat*{\section}{\songti\zihao{4}\bfseries}
\titleformat*{\subsection}{\heiti\zihao{5}\bfseries}
\titleformat*{\subsubsection}{\kaishu\zihao{5}\bfseries}

\newenvironment{csmtAbstract}{\noindent \kaishu \small {\bfseries Abstract:}}{}
\newenvironment{keywords}{\small \noindent{\bfseries Key Words:}}{}

\renewcommand{\tablename}{Tab.}

\newcommand*{\affmark}[1][*]{\textsuperscript{#1}}

\makeatletter
\renewcommand\@maketitle{%
	\hfill
	\begin{minipage}{0.95\textwidth}
		\vskip 2em
		\let\footnote\thanks 
		{\centering \bfseries \zihao{3} \@title \par} 
		\vskip 3em
		{\centering \zihao{5} \textbf{\@author} \par}
	\end{minipage}
	\vskip 1em \par
}
\makeatother

\title{A Holistic Evaluation of Piano Sound Quality}

\author{
	\footnotesize
	\textmd{\large Monan Zhou\affmark[1],  Shangda Wu\affmark[1], Shaohua Ji\affmark[1], Zijin Li\affmark[1], Wei Li\affmark[2,3\footnote{Corresponding Author: Wei Li (1970-), Male, Professor, weili-fudan@fudan.edu.cn}]}\\\vspace{0.27cm}
	\textmd{\affmark[1] Department of Music AI and Information Technology, Central Conservatory of Music, Beijing, China}\\
	\textmd{\affmark[2] School of Computer Science and Technology, Fudan University, Shanghai, China}\\\vspace{-0.17cm}
	\textmd{\affmark[3] Shanghai Key Laboratory of Intelligent Information Processing, Fudan University, Shanghai, China}
}

\begin{document}

\maketitle
\vspace{0.67cm}
\begin{csmtAbstract}
	This paper aims to develop a holistic evaluation method for piano sound quality to assist in purchasing decisions. Unlike previous studies that focused on the effect of piano performance techniques on sound quality, this study evaluates the inherent sound quality of different pianos. To derive quality evaluation systems, the study uses subjective questionnaires based on a piano sound quality dataset. The method selects the optimal piano classification models by comparing the fine-tuning results of different pre-training models of convolutional neural network (CNN). To improve the interpretability of the models, the study applies equivalent rectangular bandwidth (ERB) analysis. The results reveal that musically trained individuals are better able to distinguish between the sound quality differences of different pianos. The best fine-tuned CNN pre-trained backbone achieves a high accuracy of 98.3\% as the piano classifier. However, the dataset is limited, and the audio is sliced to increase its quantity, resulting in a lack of diversity and balance, so we use focal loss to reduce the impact of data imbalance. To optimize the method, the dataset will be expanded, or few-shot learning techniques will be employed in future research.
\end{csmtAbstract}


\section{Introduction}
\noindent
The sound quality of a piano is of paramount importance for beginners who are looking to purchase one. However, assessing sound quality accurately can be difficult for those without musical training. To aid novice piano buyers, this paper presents a method for evaluating piano sound quality using a convolutional neural network (CNN) classifier that is incorporated into a mobile application. The proposed method analyzes sound recordings of struck piano keys and generates a sound-quality score, which can help users make more informed purchasing decisions. Moreover, the method can estimate piano prices based on sound quality scores, thereby offering users more reasonable recommendations. Thus, this study has practical implications for improving the piano purchasing process for novices.

Individuals with musical training have a heightened sensitivity to the quality characteristics of piano sounds, which enables them to proficiently assess and appreciate them. Our objective is to integrate their evaluation abilities into a recognition method. To accomplish this, we first extract the spectrum of a piano recording using signal processing techniques. The resulting spectrum is then fed into a pre-trained CNN classifier, which generates probabilities of the audio belonging to each category. We subsequently calculate an expected score by combining the subjective rating conclusion with the probabilities generated by the CNN classifier. To obtain the subjective rating conclusion in the aforementioned process, we conducted a subjective questionnaire survey to obtain sound quality ratings of different pianos from individuals with musical backgrounds. We then conducted an equivalent rectangular bandwidth (ERB) analysis to determine the feasibility of differentiating sound quality and sound zones from a psychoacoustic perspective. For fine-tuning the pre-trained CNN, we first used audio slicing to increase the data size. However, this method led to a sampling imbalance. Therefore, we applied the focal loss method to reduce the impact of sample imbalance.

The existing literature on instrument sound quality has primarily focused on evaluating the sound quality of violins and pianos. For instance, in 2015, Buccoli et al. proposed an unsupervised approach \cite{buccoli2015unsupervised} for describing the semantic features of violin sound quality, while Park et al. focused on enhancing the playing techniques of violin-playing robots \cite{park2015study, jo2015study}. Suzuki investigated the impact of playing techniques on piano sound quality evaluation using spectrum analysis \cite{Suzuki2007SpectrumAA}. Similarly, Goeb et al. explored the impact of touch modes, and visual and tactile patterns on the production and perception of musical expression on the piano \cite{Goebl2014PerceptionOT}, primarily focusing on the effect of playing on sound quality. However, there is a lack of studies that assess the inherent tone quality of pianos from both subjective and objective perspectives. Therefore, this study aims to bridge this gap by developing evaluation methods for piano sound quality in the applied field.

We propose a piano sound quality evaluation method that utilizes a fine-tuned CNN pre-trained model originally designed for computer vision (CV) tasks as its backbone network. To handle audio data, we transform it into images through a short-time Fourier transform (STFT) that generates a mel spectrogram as the model input. By doing so, we can approach the audio classification problem as a CV problem. This approach has been validated by previous studies. Palanisamy et al. achieved state-of-the-art results in classifying music genres, environmental sounds, and urban sounds using fine-tuned ImageNet\footnote{\href{https://image-net.org}{https://image-net.org}}-pre-trained deep CNN models with mel spectrograms \cite{Palanisamy2020RethinkingCM}. Tsalera et al. set new records in audio classification through transfer learning \cite{Tsalera2021ComparisonOP}. Di Maggio extended this method to intelligent fault diagnosis of industrial bearings in the industrial domain \cite{DiMaggio2022IntelligentFD}. These examples demonstrate the method's high versatility in different application scenarios. Therefore, we aim to extend it to piano sound quality classification. Fortunately, the experimental results indicate that the majority of pre-trained CNN models can be adapted for piano sound quality classification tasks through fine-tuning and achieve excellent accuracy. Some lightweight pre-trained CNN models have both high accuracy and low training time costs, making them ideal choices for deployment on mobile applications.

This paper examines a dataset used for piano sound classification and presents the overall process, data processing details, and optimization method used to train the core network to address data issues. To enhance model interpretability, an ERB analysis method is employed. Additionally, a subjective questionnaire survey was conducted to evaluate piano sound quality. Performance comparison experiments of core network models for piano classification tasks are presented. Results from ERB analysis experiments are discussed, and overall conclusions are drawn.

\section{Dataset}
\noindent
To conduct the questionnaire survey and comparative experiments, a substantial amount of data is required. This part presents the dataset utilized in the experiments, along with a description of the data pre-processing procedures. The dataset and pre-processing methods were essential for conducting the experiments.

\subsection{Piano Sound Quality Dataset}
\noindent
The sound sources used in this study were obtained from the \textit{CCMUSIC}\footnote{\href{https://huggingface.co/ccmusic-database
	}{https://huggingface.co/ccmusic-database}} platform, a music data-sharing platform in China designed for musicology research. The experiment used a subset of the multifunctional music database, known as the ``Piano Sound Quality Dataset", whose first seven piano classes consist of 12 audio files (in .wav/.mp3/.m4a format) representing seven types of pianos located in the Chinese Conservatory of Music's piano rooms. These pianos are the Pearl River upright piano (PearlRiver), YOUNG CHANG upright piano (YoungChang), Steinway theater piano (Steinway-T), HSINGHAI upright piano (Hsinghai), KAWAI upright piano (Kawai), Steinway piano (Steinway), and KAWAI grand piano (Kawai-G). The dataset also contains 1,320 single-tone audio files (in .wav/.mp3/.m4a format), resulting in a total of 1,332 files. Moreover, the dataset includes a ``Subjective Questionnaire for Evaluating Piano Sound Quality" (in .xls format) that presents scoring data from 29 individuals who participated in the subjective evaluation of piano sound quality. However, we did not use the original questionnaire form and results in the dataset, but instead conducted our own questionnaire. Table \ref{tab:dataset_details} provides details of the dataset in use.

\begin{table}[H]
	\centering
	\renewcommand{\tablename}{Table}
	\caption{List of details for using the piano sound quality dataset}
	\begin{tabular}{@{}lllc@{}}
		\toprule
		\textbf{Piano name} & \textbf{Label} & \textbf{Audio}                        & \textbf{Duration} \\ \midrule
		Pearl River         & PearlRiver     & 88-note WAV file                      & 03:18             \\
		YOUNG CHANG         & YoungChang     & 88-note WAV file                      & 11:09             \\
		Steinway-Theater    & Steinway-T     & 88-note WAV file                      & 11:29             \\
		HSINGHAI            & Hsinghai       & 88-note WAV file                      & 06:42             \\
		KAWAI               & Kawai          & 88-note WAV file                      & 04:41             \\
		Steinway            & Steinway       & 52-note WAV file (missing black keys) & 04:38             \\
		KAWAI-Grand         & Kawai-G        & 88-note WAV file                      & 10:00             \\ \bottomrule
	\end{tabular}
	\label{tab:dataset_details}
\end{table}

The labels in Table \ref{tab:dataset_details} are original label names in the \textit{CCMUSIC} database. In this paper, we use piano names to represent them. Besides, all single-tone recordings of these pianos in the dataset were recorded using a uniform strength of normal pressing, eliminating the influence of playing techniques on sound quality evaluation.

\subsection{Pre-processing}
\noindent
In some past \textit{Kaggle}\footnote{\href{https://www.kaggle.com
	}{https://www.kaggle.com}} competitions for CV, the size of a pre-trained CNN, the number of classification categories, and the amount of data are typically related in a ratio of approximately $1:2k$ based on past experience. For a seven-classification problem, it is recommended to control the sample size at around $14k$. However, when evaluating the sound quality of piano recordings, using each recording as a sample may not provide adequate data. To address the issue of insufficient data, data augmentation methods were considered. Traditional image data augmentation methods such as mixup \cite{mixup}, AugMix \cite{augmix}, Cutout \cite{cutout}, Mosaic \cite{yolov4}, and CutMix \cite{Yun2019CutMixRS} cannot be applied to this task because they may damage the original sound quality information. Instead, this study uses a sliding window approach to slice audio segments and convert them into mel spectrograms to extract features. The sliding window has a fixed length of 0.2 seconds with a step length of 0.2 seconds. The function used to convert audio to mel spectrogram is from the \textit{librosa}\footnote{\href{https://pypi.org/project/librosa}{https://pypi.org/project/librosa}} library, with a sampling rate of n\_fft set to 1,024. Since the time-frequency window is narrow, setting the default parameter n\_fft=2,048 would result in window overflow errors in the program.

To comply with the recommended amount of data for a seven-classification task, a step length of 0.2 seconds was used for data augmentation, resulting in exactly 14,415 data points. To maintain data independence and consistency, the time-frequency window was also set to 0.2 seconds. Setting the time-frequency window width greater than the step length results in a high degree of overlap in the spectrogram of adjacent frames, rendering labeled data essentially the same. High overlap in training and test sets may lead to decreased distinguishability between the two sets, resulting in highly overlapping validation and training accuracy curves that fail to provide effective verification.

\section{Methodology}
\noindent
This section outlines key methods used in our paper. It begins by describing our overall process, followed by an explanation of ERB's application for interpreting the piano timbre classification task from a perceptual acoustics perspective. Lastly, it introduces the use of focal loss to address sample imbalance issues after data processing in the comparative experiment.

\subsection{Overall Process}
\noindent
The whole process of our evaluation method is illustrated in Figure \ref{fig:workflow}:

\begin{figure}[H]
	\vspace{-5em}
	\centering
	\includegraphics[scale=0.45]{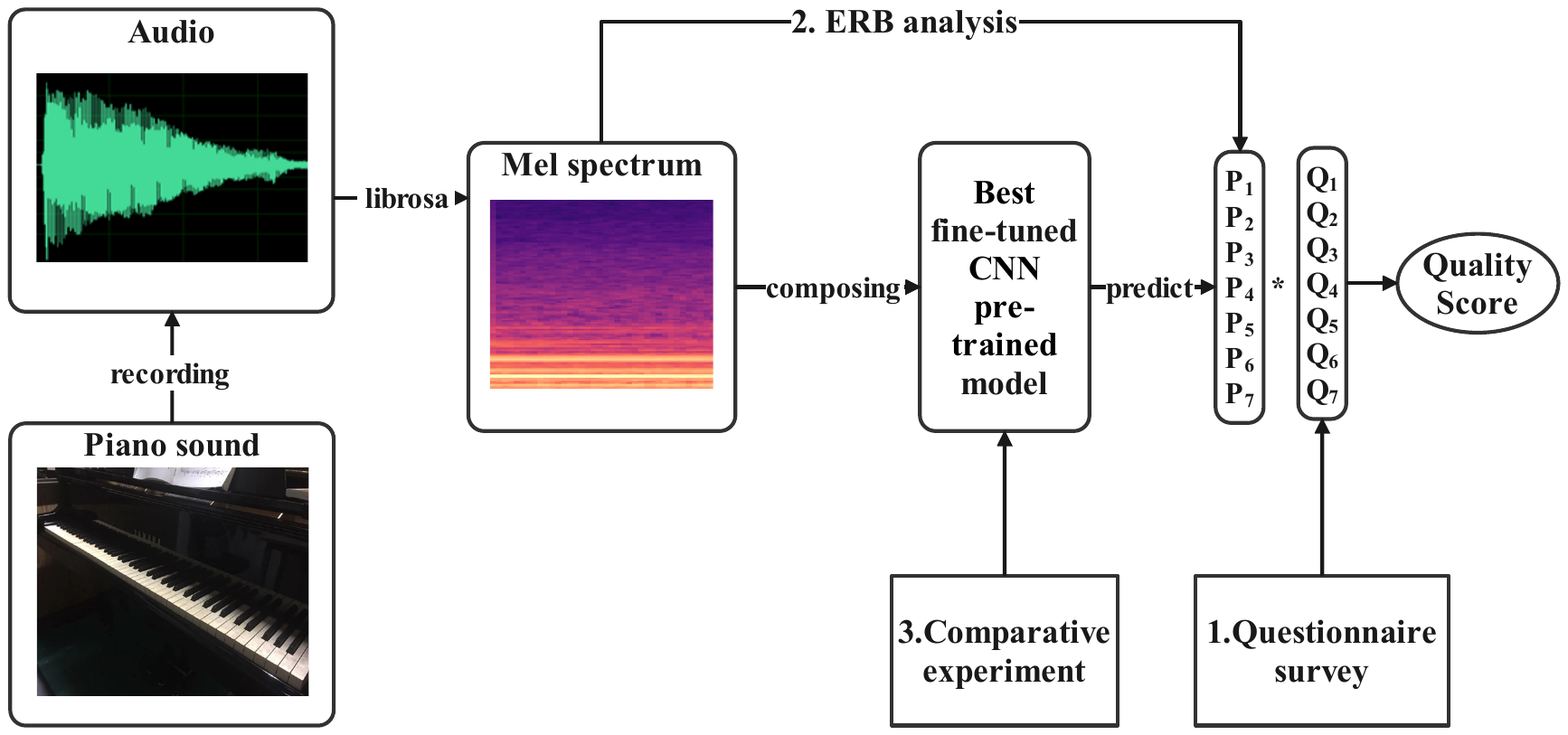}
	\vspace{-4em}
	\caption{The overall process of our method}
	\label{fig:workflow}
\end{figure}

In Figure \ref{fig:workflow}, $P_1 - P_7$ represent the probabilities predicted by the CNN model that the audio recording belongs to each of the seven types of pianos, while $Q_1 - Q_7$ represent the sound quality of the seven types of pianos. Our contributions can be summarized as follows:

\begin{enumerate}
	\item We obtained sound quality scores for the dataset by conducting a subjective experiment with professional pianists.
	\item We used the ERB method to analyze the frequency distribution of each piano type in the dataset. This provided a white-box interpretation of how the model distinguishes between different piano sound qualities based on the frequency spectrum.
	\item We compared pre-trained CNN backbones to find the best classifier for this application scenario.
\end{enumerate}

The method illustrated in Figure \ref{fig:workflow} separates the classifier from the evaluation method, allowing for easy maintenance by updating the evaluation method as subjective evaluation systems change. The use of a classification task rather than regression lowers the dataset's quality threshold, as it reduces the emphasis on recording equipment, environment, and techniques. However, this approach does not provide a comprehensive understanding of the true distribution of piano sound quality in the acoustic space. Instead, it only offers a rough estimate of sound quality grade based on piano brands, which is still a useful reference for beginners to avoid making uninformed purchases. Adding more piano types to the model will improve its grasp of the distribution, resulting in more accurate predictions.

\subsection{ERB Representation}
\noindent
The ERB is a psychoacoustic technique used to model the auditory filter of the human ear. The ERB filter's bandwidth is equivalent to an ideal rectangular filter with the same transmission gain as the given filter. If white noise is input, the power transmitted through the ERB filter is equal to that of the given filter. Brian Moore proposed a polynomial equation as Eq. (\ref{eq:erb1}) for estimating the bandwidth of the auditory filter, specifically for young listeners and moderate sound levels \cite{Moore1983SuggestedFF}. This method is an essential tool for understanding how humans perceive sounds, especially in music and speech contexts.

\begin{equation}
	ERB(f) = 6.23f^2 + 93.39f + 28.52
	\label{eq:erb1}
\end{equation}

Another more accurate approximation shown as Eq. (\ref{eq:erb2}) is given in \cite{Glasberg1990DerivationOA}:

\begin{equation}
	ERB(f)=24.7(4.37f+1)
	\label{eq:erb2}
\end{equation}

In this study, we employed the \textit{TimbreToolbox} \cite{Peeters2011TheTT} to examine the timbres in our dataset and offer a psychoacoustic interpretation for the variations in sound quality observed among participants. We used the toolbox to compute the ERB representation of each piano note, which yields a psychological description of the listener's perception of each time-frequency band in the sound signal.

We computed the ERB representation using \textit{TimbreToolbox} by applying a time-domain window of 0.0058 seconds (256 sampling points) and a frequency-domain window of 77. The bandwidth of the frequency-domain window increased exponentially from 0 to 16 kHz, enabling us to calculate the ERB and the filtered signal power spectrum within each window. To analyze the ERB, we used recordings of different durations, such as 1 second, 1.2 seconds, and the full duration of the piano note recording (approximately 2 seconds). The piano note recordings were about 2 seconds long.

To differentiate between sound quality and sound zones, we utilized principal component analysis (PCA) \cite{Harman1961ModernFA} and t-distributed stochastic neighbor embedding (t-SNE) \cite{Maaten2008VisualizingDU} methods to graph ERB representation data on a two-dimensional plane. To ensure uniformity in our dataset, we adjusted the duration of recordings, which ranged from under 1.5 seconds to roughly 2 seconds, to a standardized length of 1.2 seconds for analysis purposes.

\subsection{Focal Loss}
\noindent
The focal loss \cite{lin2017focal} is a loss function that addresses the class imbalance in deep learning models. In the context of a piano sound quality dataset, recording lengths vary across different piano ranges and models. Despite an initially balanced sample size in terms of audio quantity, the slight differences in recording length can result in significant imbalances in sample size after 0.2-second slicing. Table \ref{tab:fl} presents the distribution of the number of samples for each piano after slicing.

\begin{table}[H]
	\renewcommand{\tablename}{Table}
	\caption{Distribution of the number of piano data after slicing processing}
	\resizebox{\linewidth}{!}{
		\begin{tabular}{@{}cccccccc@{}}
			\toprule
			\textbf{Piano label} & PearlRiver & YoungChang & Steinway-T & Hsinghai & Kawai & Steinway & Kawai-G \\ \midrule
			\textbf{Support}     & 73         & 338        & 336        & 198      & 131   & 134      & 232     \\ \bottomrule
		\end{tabular}
		\label{tab:fl}
	}
\end{table}

To mitigate the effects of sample imbalance, we utilized focal loss as the loss function for CNN backpropagation. The focal loss was originally introduced to address the imbalance between positive and negative samples and was designed as an optimization technique for binary classification problems. However, our task involved multi-classification. Thus, we extended the original focal loss formula to accommodate multi-classification tasks. The resulting calculation formula is as Eq. (\ref{eq:fl}).

\begin{equation}
	FL(p_t)=-\sum_i \alpha_i {(1-p_t)}^\gamma \log(p_t)
	\label{eq:fl}
\end{equation}

The weight value $\alpha_i$ is a decimal between 0 and 1 and must conform to the zero-sum game principle, as in the original focal loss formula. In the focal loss formula, positive and negative samples have weights of $\alpha$ and $1 - \alpha$, respectively. When dealing with multiple positive samples, the weight $\alpha$ should decrease as the number of positive samples increases. In the case of multi-classification tasks, each class of samples should have its weight. For example, in a k-classification problem, the weight of the i-th class of samples is $\alpha_i$. Therefore, the weights $\alpha_i$ must satisfy the conditions in Eq. (\ref{eq:sum}).

\begin{equation}
	\sum_{i=1}^k \alpha_i = 1\quad (0\leq \alpha_i\leq 1, k\geq 2)
	\label{eq:sum}
\end{equation}

To allocate weights to classes of data based on the principle that weight decreases as sample size increases, we can use the inverse of the sample size. We denote the sample size of the k-th class of data as $s_k$. The weight calculation for k classes of data can be expressed mathematically as follows:

\begin{equation}
	[\alpha_1, \alpha_2, ..., \alpha_k] = [s_1^{-1}, s_2^{-1}, ..., s_k^{-1}] / \sum_{i=1}^k s_i^{-1} \quad(k\geq 2)
	\label{eq:alpha}
\end{equation}

In the context of our task, the weights assigned to the seven piano types upon calculation are shown in Eq. (\ref{eq:alphas}).

\begin{equation}
	[\alpha_1, \alpha_2, ..., \alpha_7]=[0.3182, 0.0673, 0.0663, 0.1128, 0.1605, 0.1618, 0.1131]
	\label{eq:alphas}
\end{equation}

The weight array in the dataset represents the current data distribution. Any changes in the quantity or category distribution of the dataset will trigger our program to re-assign weights based on the actual data distribution, using the formula (\ref{eq:alpha}). The value of $\gamma$ in the extended focal loss formula becomes 0, which means that the focal loss degenerates to a weighted cross-entropy loss.

\section{Experiment}
\noindent
We administered a questionnaire survey to individuals with musical backgrounds to obtain quality ratings for various pianos, and with ERB analysis to enhance method interpretability, conducted a comparative experiment on piano classification using different pre-trained CNNs.

\subsection{Questionnaire Survey}
\noindent
The aim of this study was to develop a neural network capable of accurately mimicking human perception of piano sound quality. To achieve this, we evaluated piano sound quality across the entire range of 88 keys, with attention to the low, middle, and high registers. The study involved 30 piano performance majors from the Xinghai Conservatory of Music who listened to seven piano recordings, covering all 88 keys and rated the sound quality of the low, middle, and high registers. To minimize potential bias from brand recognition, we conducted a blind listening test in which participants were unaware of the piano's brand during the test. This was done to ensure impartial evaluations.

To facilitate sound quality assessments across different ranges, we provided rating prompts to participants. For the low range, we asked them to identify deep and rich sounds while avoiding any muffled or unclear audio. The middle range was assessed for suitable loudness and comfort, while the high range was rated for full and smooth sound without any sharp or broken notes. Participants used a 5-point rating scale, with 1 indicating very poor and 5 indicating excellent, to rate each piano recording.

\begin{figure}[H]
	\centering
	\includegraphics[scale=0.8]{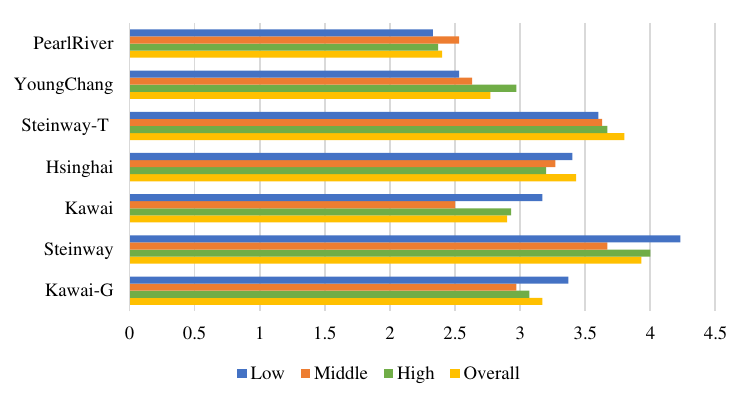}
	\vspace{-1em}
	\caption{Sound quality ratings of 7 piano recordings by 30 piano performance students for low, middle, high, and overall ranges, given on a scale of 1 to 5, with 1 being ``very poor" and 5 being ``excellent"}
	\label{fig:survey1}
\end{figure}

Figure \ref{fig:survey1} displays significant variations in sound quality ratings across different piano brands. Steinway received the highest rating of 3.93, followed by Steinway-T at 3.8, based on the overall score obtained from seven piano recordings. In contrast, PearlRiver had the lowest overall rating of 2.4, with YoungChang following closely at 2.77. These findings suggest that professionals perceive significant differences in sound quality between piano brands.

\begin{figure}[H]
	\centering
	\includegraphics[scale=0.8]{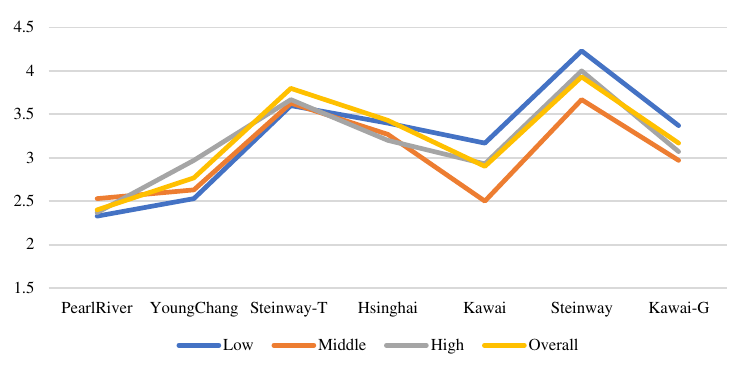}
	\vspace{-1em}
	\caption{Correlation between sound quality ratings for low, middle, high, and overall ranges of 7 piano recordings, with correlation coefficients calculated based on the ratings given by 30 piano performance students}
	\label{fig:survey2}
\end{figure}

Figure \ref{fig:survey2} displays the correlation coefficients of ratings assigned to seven piano recordings, categorized into low, middle, high, and overall ranges. The correlation coefficient measures the strength of the linear relationship between two variables and ranges from -1 to 1. A coefficient of 1 indicates a perfect positive correlation, while a coefficient of -1 indicates a perfect negative correlation. A coefficient of 0 indicates no correlation. Eq. (\ref{eq:corr}) can be used to calculate the correlation coefficient.

\begin{equation}
	Corr(X,Y)=\frac{\sum\limits_{i=1}^{n}(x_i-\overline{x})(y_i-\overline{y})}{\sqrt{\sum\limits_{i=1}^{n}(x_i-\overline{x})^2\sum\limits_{i=1}^{n}(y_i-\overline{y})^2}}
	\label{eq:corr}
\end{equation}

Strong positive correlations were found between sound quality ratings for the low, middle, and high ranges and the overall sound quality rating, with correlation coefficients of 0.9411, 0.9596, and 0.9669, respectively. These results indicate consistent sound quality ratings across the different ranges, suggesting no significant difference in sound quality perception between the low, middle, and high ranges. The findings imply uniform sound quality across the frequency spectrum.

Our objective was to transfer human perception of piano sound quality to a neural network by utilizing the sound quality evaluation system created by individuals with musical backgrounds as the weighted classification results.

\subsection{ERB Analysis}
\begin{figure}[H]
	\centering
	\begin{minipage}[b]{0.31\linewidth}
		\includegraphics[width=\linewidth]{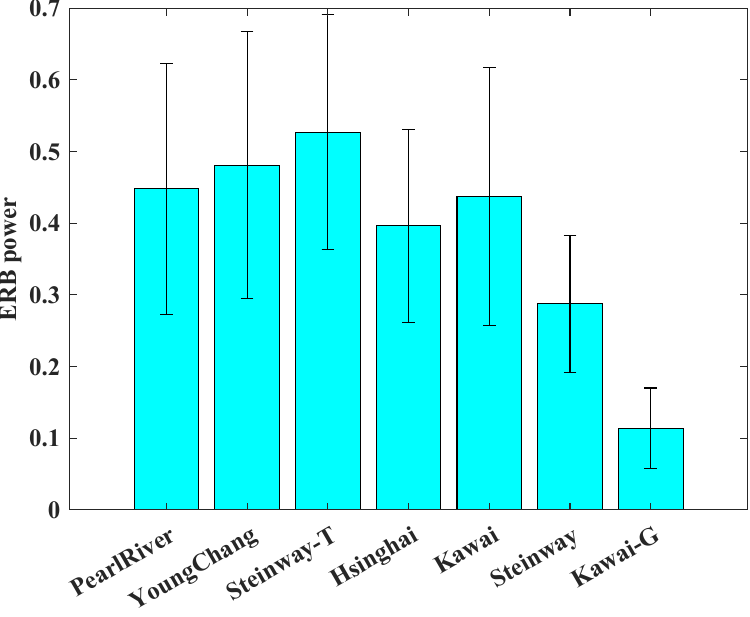}
		\vspace{-2em}
		\caption*{(a)}
	\end{minipage}
	\hfill
	\begin{minipage}[b]{0.31\linewidth}
		\includegraphics[width=\linewidth]{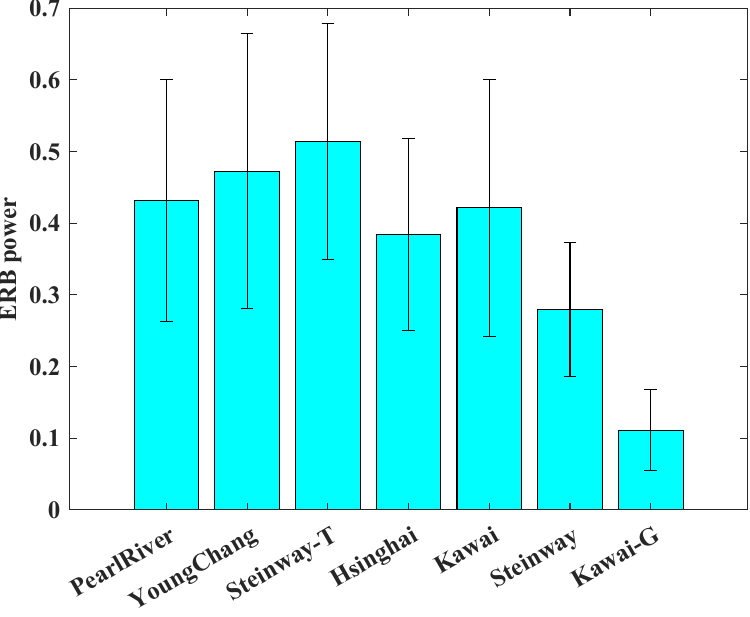}
		\vspace{-2em}
		\caption*{(b)}
	\end{minipage}
	\hfill
	\begin{minipage}[b]{0.31\linewidth}
		\includegraphics[width=\linewidth]{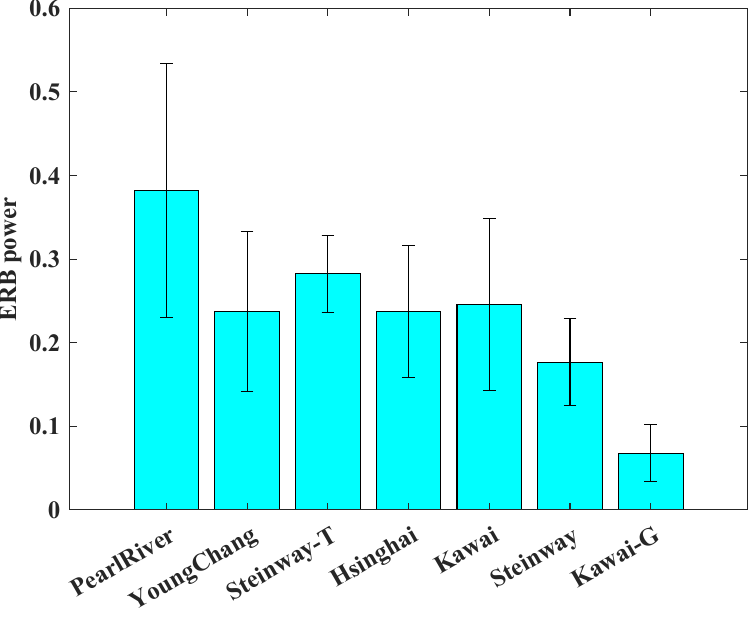}
		\vspace{-2em}
		\caption*{(c)}
	\end{minipage}
	\vspace{-1em}
	\caption{Average ERBs for each piano brand in (a) 1s duration, (b) 1.2s duration, and (c) whole duration}
	\label{fig:erb_avg}
\end{figure}

\noindent
To gain insights into the interpretability of the piano timbre classification task from the perspective of perceptual acoustics, we analyzed the ERB of piano timbre. Figure \ref{fig:erb_avg} displays the average ERB representation for each piano brand.

\begin{figure}[H]
	\centering
	\begin{minipage}[b]{0.31\linewidth}
		\includegraphics[width=\linewidth]{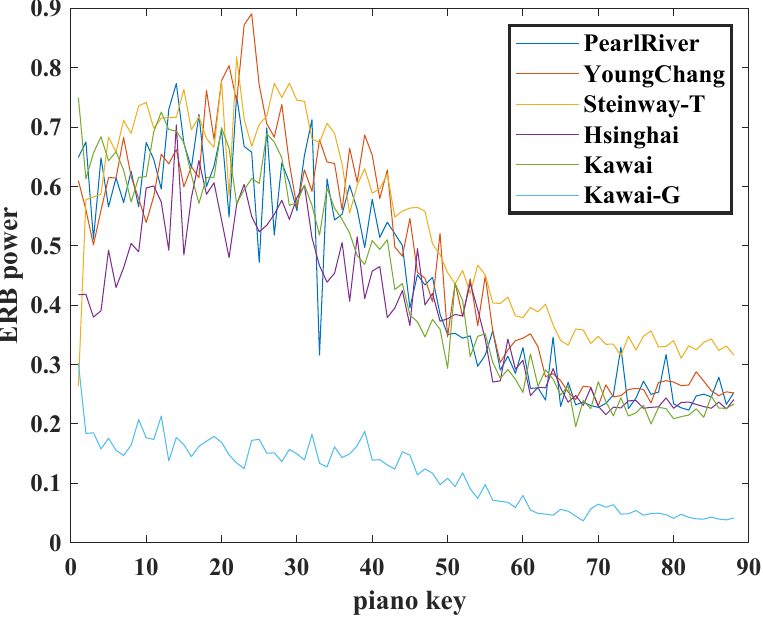}
		\vspace{-2em}
		\caption*{(a)}
	\end{minipage}
	\hfill
	\begin{minipage}[b]{0.31\linewidth}
		\includegraphics[width=\linewidth]{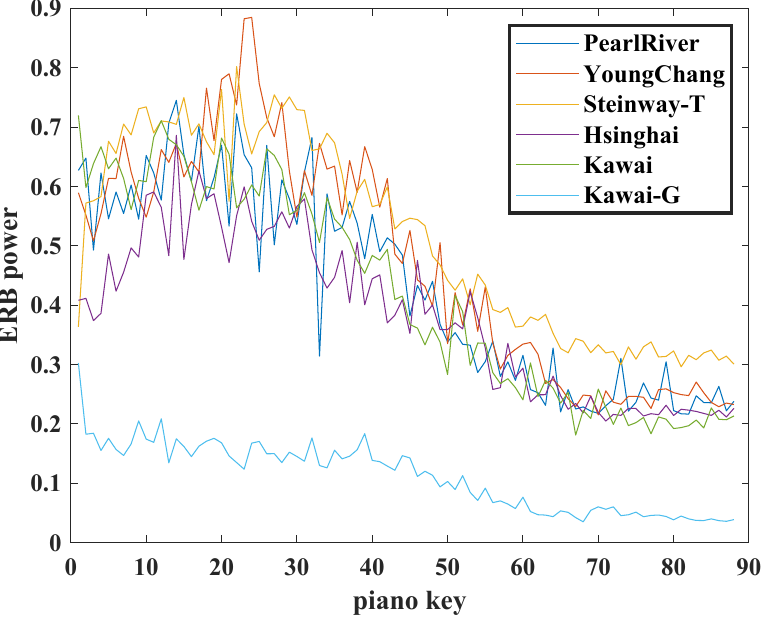}
		\vspace{-2em}
		\caption*{(b)}
	\end{minipage}
	\hfill
	\begin{minipage}[b]{0.31\linewidth}
		\includegraphics[width=\linewidth]{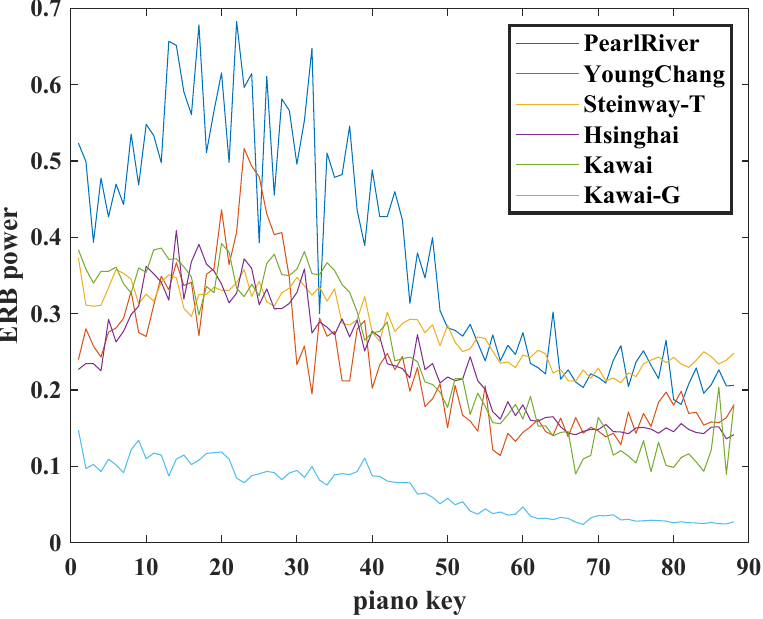}
		\vspace{-2em}
		\caption*{(c)}
	\end{minipage}
	\vspace{-1em}
	\caption{Mean ERBs for each piano brand in (a) 1s duration, (b) 1.2s duration, and (c) whole duration}
	\label{fig:mean_erb}
\end{figure}

Figure \ref{fig:erb_avg} reveals that the dataset's varying recording times lead to differences in results between analyses of the entire recording length and a specific length. However, the discrepancy between results obtained from analyzing 1s and 1.2s of the recording is negligible. Typically, pianos with inferior sound quality exhibit a larger average ERB representation, except for Steinway, which has only 52 tones in the dataset. This discrepancy may also be due to variations in recording quality. Figure \ref{fig:mean_erb} presents the calculated results of the single-note ERB representation for each pitch of each piano.

The average ERB representation values for all pianos decrease as the pitch increases, peaking in the lower and middle registers. This trend reflects the fact that higher frequency bands correspond to smaller ERB bandwidths and, therefore, smaller representation values from a perceptual acoustics standpoint. Moreover, pianos with superior sound quality exhibit smoother average ERB curves, which can aid in interpreting piano timbre quality from a perceptual perspective.

\begin{figure}[H]
	\centering
	\begin{minipage}[b]{0.48\linewidth}
		\includegraphics[width=\linewidth]{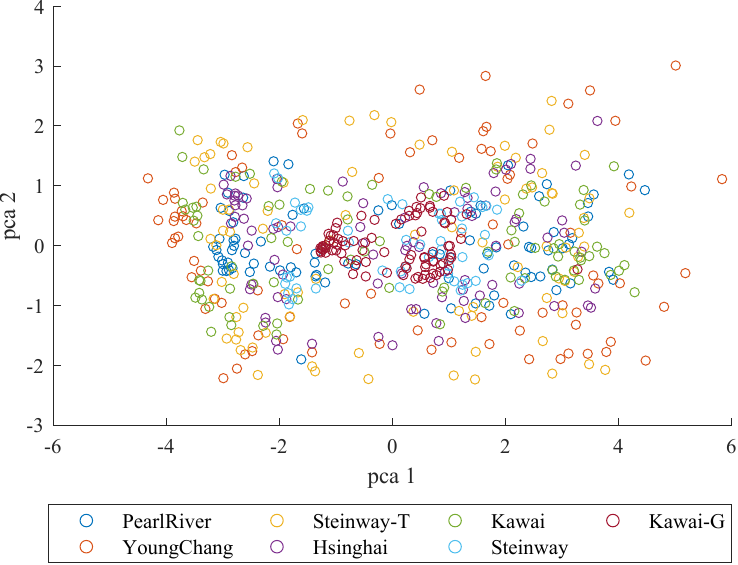}
		\vspace{-2em}
		\caption*{(a)}
	\end{minipage}
	\hfill
	\begin{minipage}[b]{0.48\linewidth}
		\includegraphics[width=\linewidth]{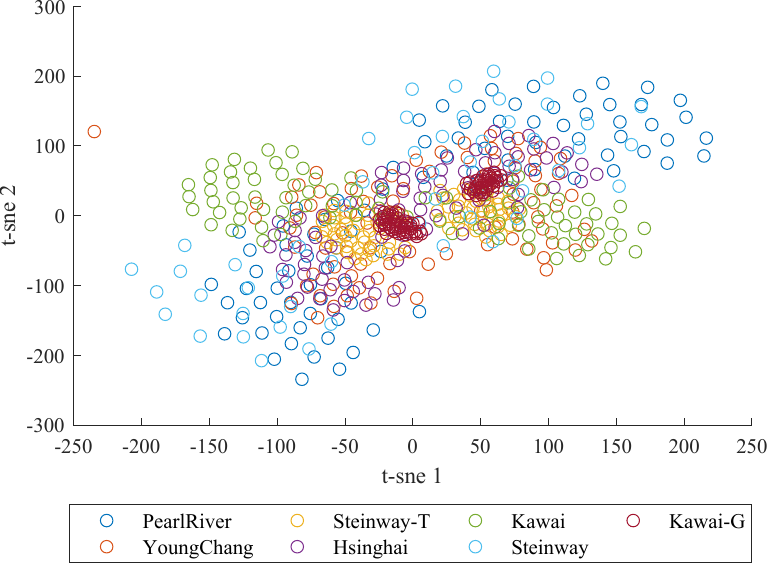}
		\vspace{-2em}
		\caption*{(b)}
	\end{minipage}
	\vspace{-1em}
	\caption{PCA (a) and t-SNE (b) visualizations of different brands of pianos in 1.2s duration}
	\label{fig:pca_tsne}
\end{figure}

We employed PCA and t-SNE algorithms to visualize the ERB representation for sound quality discrimination, using only the 1.2s recording duration for analysis. Figure \ref{fig:pca_tsne} presents the corresponding results for further clarification.

\begin{figure}[H]
	\centering
	\begin{minipage}[b]{0.24\linewidth}
		\includegraphics[width=\linewidth]{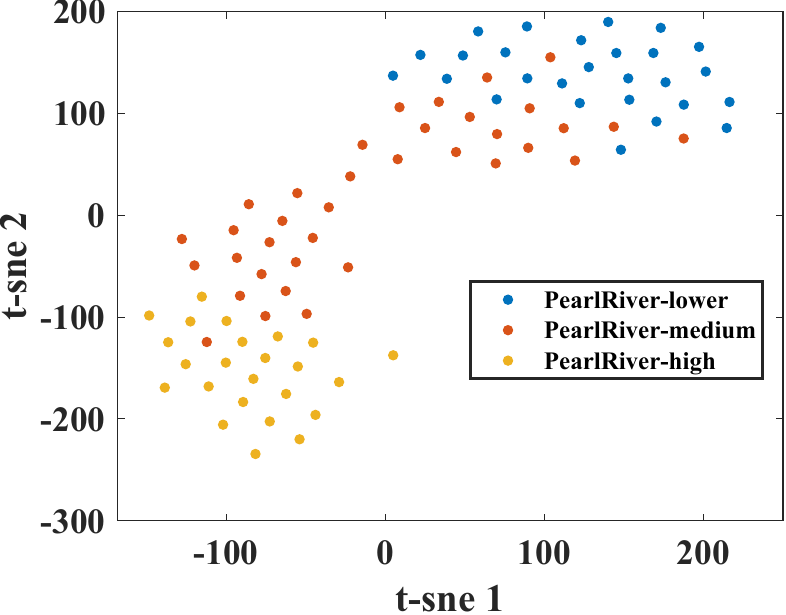}
		\vspace{-2em}
		\caption*{(a)}
	\end{minipage}
	\hfill
	\begin{minipage}[b]{0.24\linewidth}
		\includegraphics[width=\linewidth]{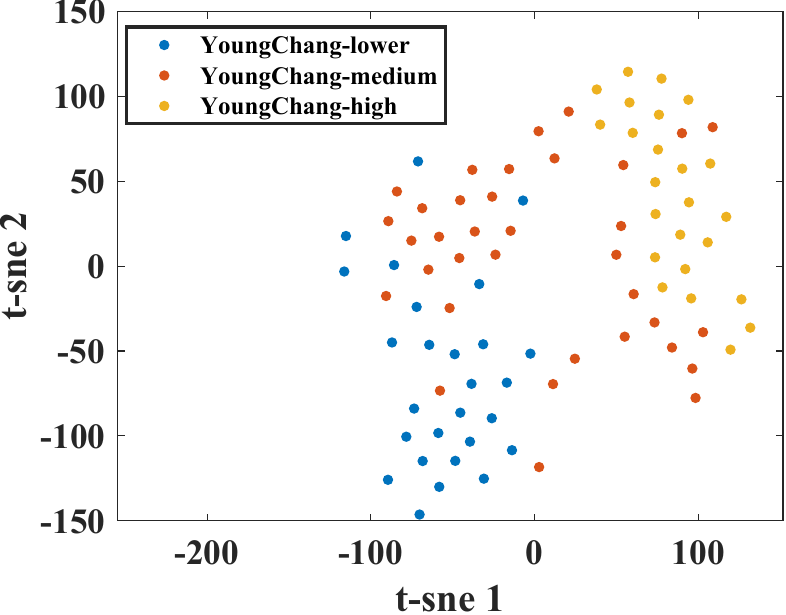}
		\vspace{-2em}
		\caption*{(b)}
	\end{minipage}
	\hfill
	\begin{minipage}[b]{0.24\linewidth}
		\centering
		\includegraphics[width=\linewidth]{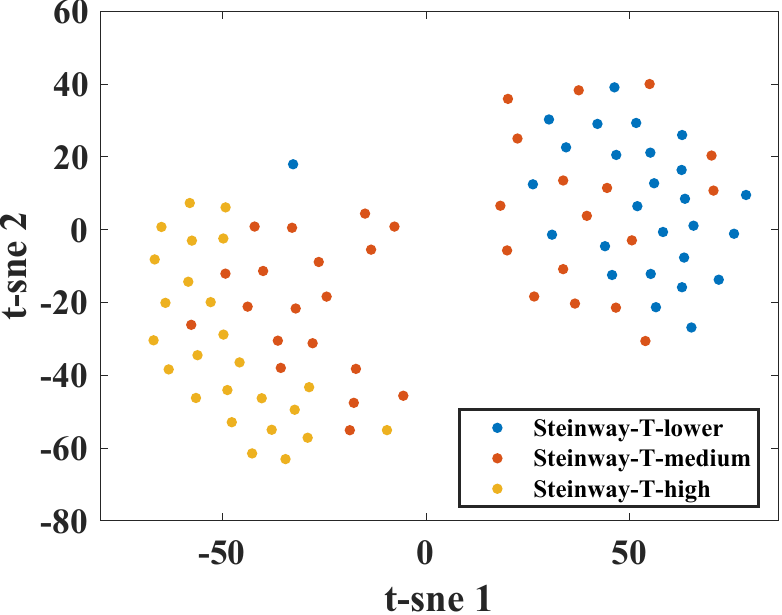}
		\vspace{-2em}
		\caption*{(c)}
	\end{minipage}
	\hfill
	\begin{minipage}[b]{0.24\linewidth}
		\includegraphics[width=\linewidth]{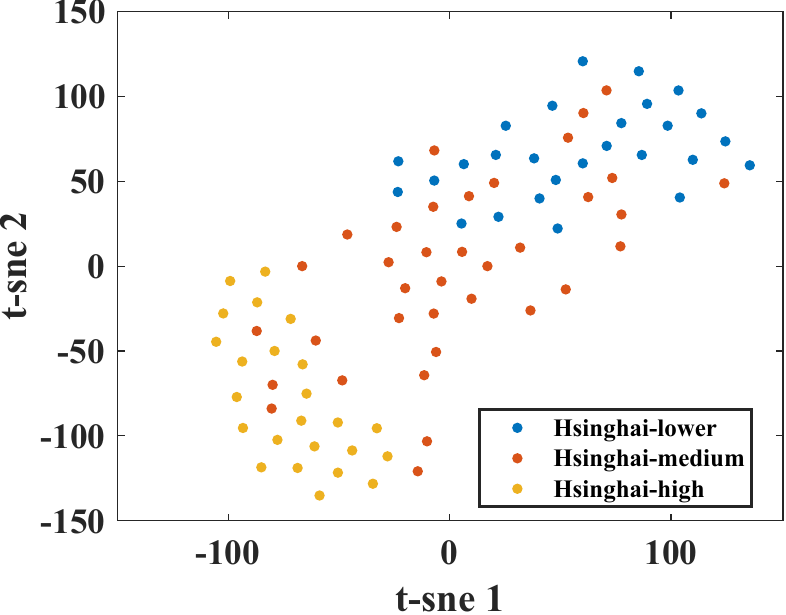}
		\vspace{-2em}
		\caption*{(d)}
	\end{minipage}

	\begin{minipage}[b]{0.24\linewidth}
		\centering
		\includegraphics[width=\linewidth]{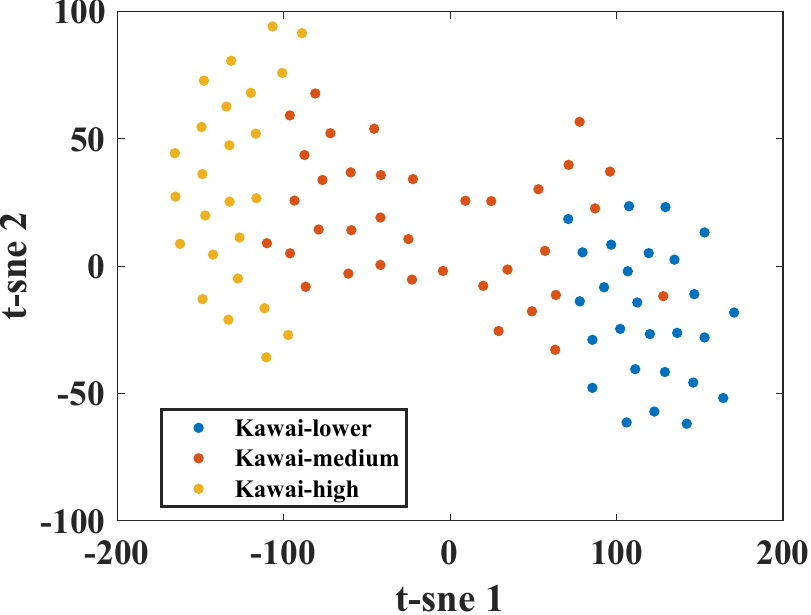}
		\vspace{-2em}
		\caption*{(e)}
	\end{minipage}
	\begin{minipage}[b]{0.24\linewidth}
		\centering
		\includegraphics[width=\linewidth]{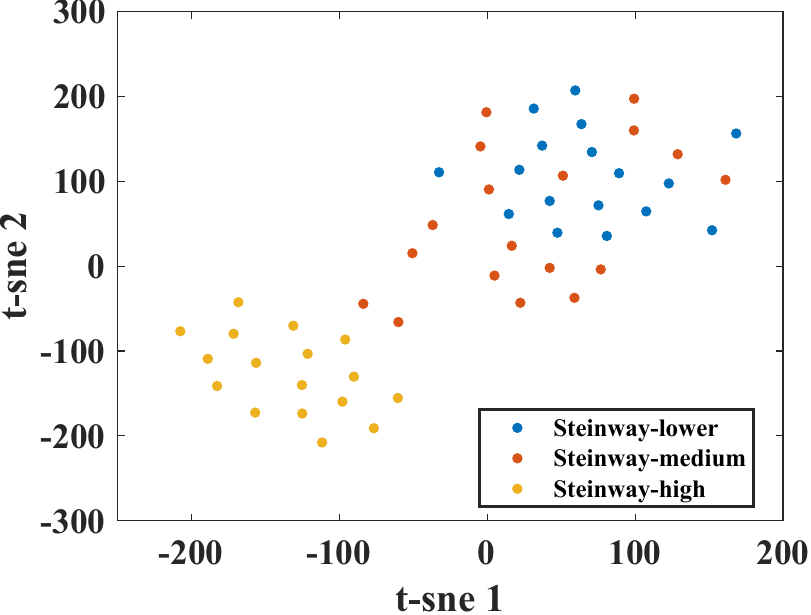}
		\vspace{-2em}
		\caption*{(f)}
	\end{minipage}
	\begin{minipage}[b]{0.24\linewidth}
		\centering
		\includegraphics[width=\linewidth]{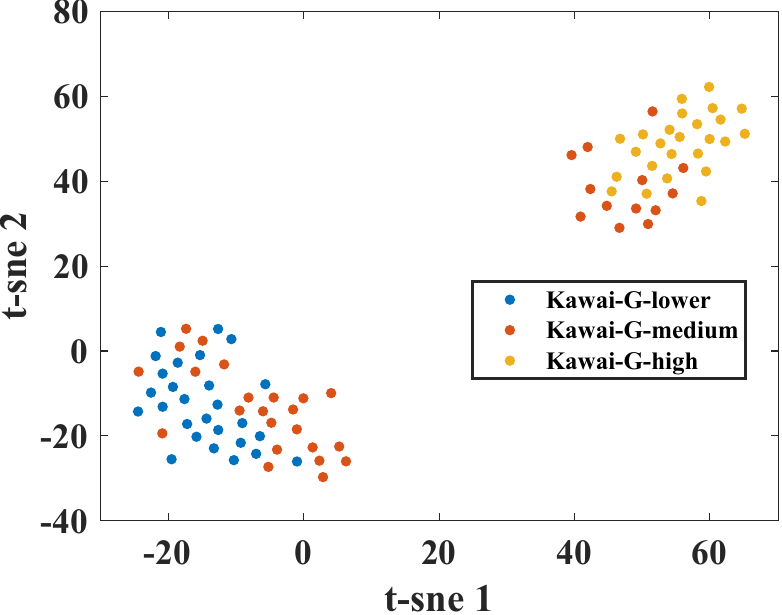}
		\vspace{-2em}
		\caption*{(g)}
	\end{minipage}
	\vspace{-1em}
	\caption{t-SNE visualizations of the three pitch regions for (a) PearlRiver, (b) YoungChang, (c) Steinway-T, (d) Hsinghai, (e) Kawai, (f) Steinway, and (g) Kawai-G}
	\label{fig:tsne_vi}
\end{figure}

The visualization results were intriguing, as it was challenging to observe all pitches produced by a single piano clustering together on the two-dimensional PCA representation, although Kawai-G and Steinway exhibited noticeable clustering. Conversely, using the t-SNE manifold embedding method, most pitches generated by different pianos could be clustered into one or two groups. Based on these observations, we conducted t-SNE visualizations of the three pitch regions for each piano, as depicted in Figure \ref{fig:tsne_vi}.

\subsection{Comparative Experiment}
\noindent
The purpose of this comparative experiment was to select the most advanced architecture from each backbone in the field of CV. To fine-tune the models, we compared the accuracy and training cost of the linear probing mode, which updates only the weight parameters of the classifier near the output layer, with the full fine-tuning mode, which updates all trainable weight parameters, including those in the classifier and backbone, by backpropagation. The latter mode increases training time costs.

\begin{figure}[H]
	\centering
	\includegraphics[scale=0.45]{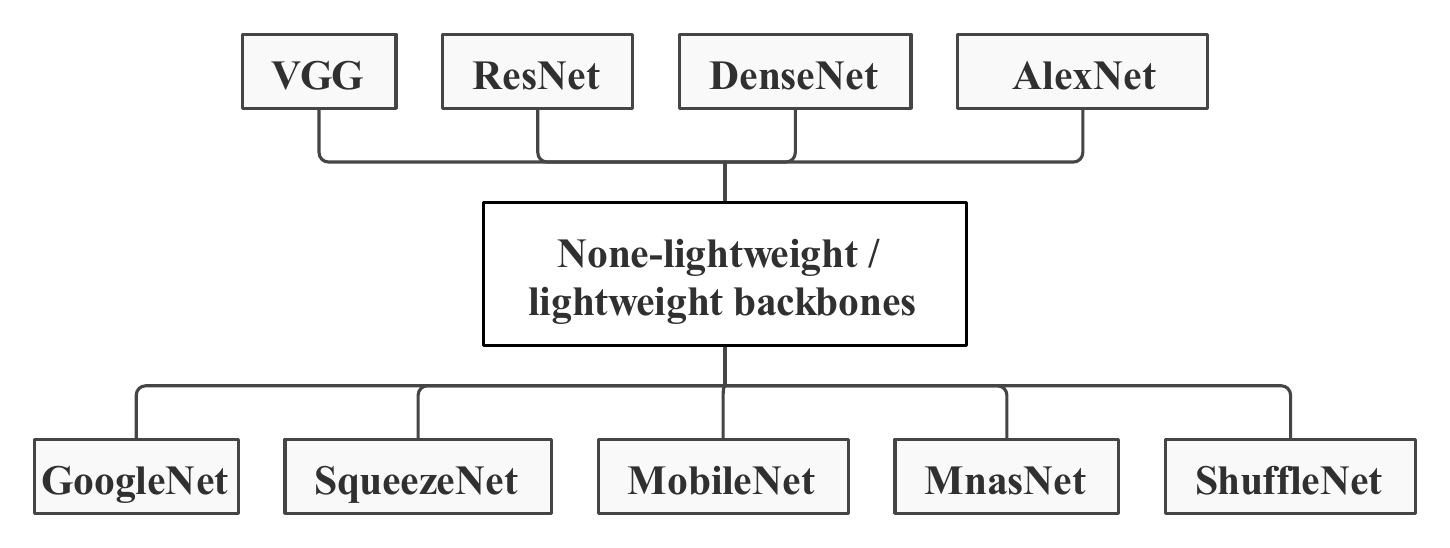}
	\caption{Backbone types used in our comparative experiment}
	\label{fig:mode_backbone}
\end{figure}

Figure \ref{fig:mode_backbone} displays the backbone types utilized in our experiment, which includes both relatively large CNN models, such as VGG \cite{Simonyan2014VeryDC}, ResNet \cite{He2015DeepRL}, DenseNet \cite{Huang2016DenselyCC}, and AlexNet \cite{Krizhevsky2012ImageNetCW}, and lightweight CNN models, such as ShuffleNet \cite{Zhang2017ShuffleNetAE}, MobileNet \cite{Harjoseputro2020MobileNetsEC}, SqueezeNet \cite{Iandola2016SqueezeNetAA}, GoogleNet \cite{Szegedy2014GoingDW}, and MnasNet \cite{Tan2018MnasNetPN}, most of which are primarily designed for mobile devices. The design objectives of the lightweight models are to achieve optimal model accuracy using limited computational resources. Table \ref{tab:comparexp} presents a compiled list of comparative experimental results, where the F1-score represents the weighted average of the F1-score, and time-cost refers to the time required to complete fine-tuning.

\begin{table}[H]
	\renewcommand{\tablename}{Table}
	\caption{Comparison experiment of various backbone models in the CV field for piano sound quality classification task}
	\label{tab:comparexp}
	\resizebox{\linewidth}{!}{
		\begin{tabular}{@{}cccccc@{}}
			\toprule
			\multirow{2}{*}{\textbf{Backbone}}               & \multirow{2}{*}{\textbf{Parameter count(Mb)}} & \multicolumn{2}{c}{\textbf{Linear probing}} & \multicolumn{2}{c}{\textbf{Full fine-tuning}}                                                 \\ \cmidrule(l){3-6}
			                                                 &                                               & \textbf{Accuracy(\%)}                       & \textbf{F1-score(\%)}                         & \textbf{Accuracy(\%)} & \textbf{F1-score(\%)} \\ \midrule
			VGG19-BN \cite{Simonyan2014VeryDC}               & 143.7                                         & \textbf{83.3}                               & \textbf{83.4}                                 & 98.2                  & 98.2                  \\ \midrule
			Wide ResNet-101 V2 \cite{He2015DeepRL}           & 126.9                                         & 46.1                                        & 45.4                                          & 97.4                  & 97.4                  \\ \midrule
			AlexNet \cite{Krizhevsky2012ImageNetCW}          & 61.1                                          & 83.1                                        & 83.0                                          & 97.9                  & 97.9                  \\ \midrule
			DenseNet-201 \cite{Huang2016DenselyCC}           & 20.0                                          & 45.2                                        & 44.9                                          & 92.8                  & 92.8                  \\ \midrule
			MnasNet 1.0 \cite{Tan2018MnasNetPN}              & 4.4                                           & 44.4                                        & 44.2                                          & 79.8                  & 79.8                  \\ \midrule
			MobileNet V2 \cite{Harjoseputro2020MobileNetsEC} & 3.5                                           & 59.2                                        & 58.7                                          & 94.4                  & 94.4                  \\ \midrule
			SqueezeNet 1.1 \cite{Iandola2016SqueezeNetAA}    & 1.2                                           & 74.9                                        & 75.2                                          & \textbf{98.3}         & \textbf{98.3}         \\ \bottomrule
		\end{tabular}
	}
\end{table}

\begin{figure}[H]
	\centering
	\includegraphics[scale=0.55]{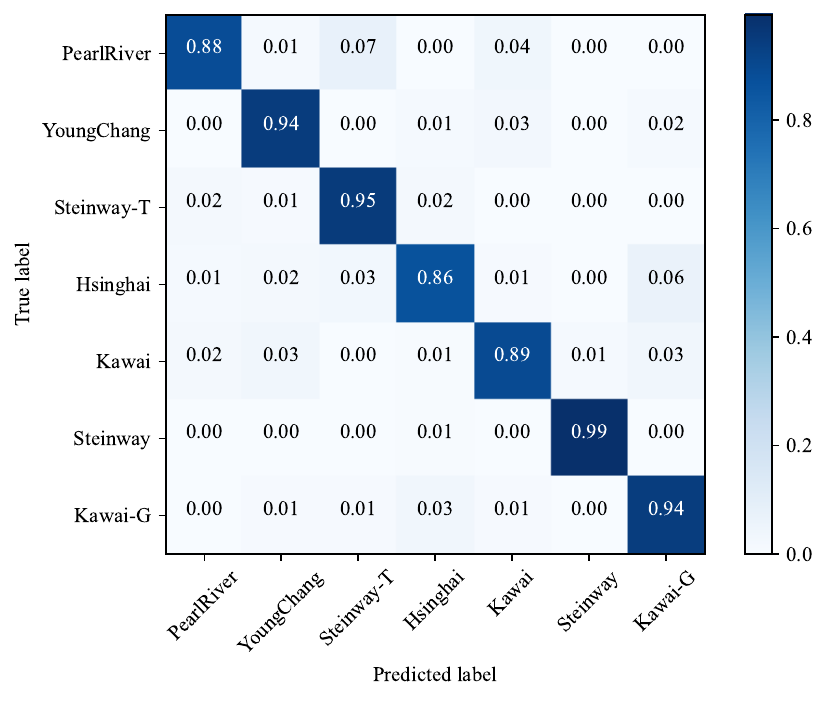}
	\caption{The confusion matrix of pre-trained SqueezeNet in full fine-tuning mode}
	\label{fig:squeezenet-cm}
	\vspace{-1em}
\end{figure}

The performance results of the GoogleNet and ShuffleNet backbone families were not included in our task's table due to the experiment's findings that their structures or pre-trained weights were not suitable for transfer learning in our specific task. Furthermore, we present Figure \ref{fig:squeezenet-cm}, which depicts the confusion matrix of SqueezeNet in full fine-tuning mode. SqueezeNet demonstrated the best performance among the backbones in our comparative experiment.

There are some advanced CV models related to music such as convolutional recurrent neural network (CRNN) \cite{Shi2015AnET} and musicnn \cite{Pons2019musicnnPC}. However, CRNN is not pre-trained by ImageNet and can be used for musical score recognition instead of mel spectrogram classification. Although musicnn is pre-trained using music information, it has a high model complexity, and fine-tuning with insufficient data can lead to overfitting. Additionally, our algorithm's original goal was to enable the deployment of a piano sound quality scoring app on mobile devices, necessitating the use of lightweight CNN networks like MobileNet and ShuffleNet. However, these networks have not been pre-trained on music datasets like musicnn. Therefore, we can only employ ImageNet-pre-trained backbones for fine-tuning within the CV domain. To ensure variables are controlled in comparison experiments against non-lightweight networks, we also utilize ImageNet-pre-trained models across all models.

\section{Conclusion}
\noindent
Listeners with musical backgrounds exhibit high subjective discrimination of piano sound quality between different brands, with discrimination scores ranging from 2.4 to 3.93. However, subjective discrimination of sound quality among different tonal regions within the same piano is low, with correlation coefficients approaching 1.0. These findings suggest that individuals with musical training possess a heightened ability to discern differences in piano sound quality across various brands, but differences in sound quality across different tonal regions within a single piano are less perceptible.

The feasibility of differentiating sound quality and sound zones from a psychoacoustic perspective was confirmed through ERB characterization analysis, which involves calculating the signal power spectrum after ERB filtering. Therefore, distinguishing sound quality through spectrograms is feasible.

The fine-tuned SqueezeNet pre-trained model was deemed more suitable for the task due to its higher accuracy and lightweight structure, making it ideal for deployment on mobile applications. By replacing its softmax function with normalized probabilities in the header, which can then be used with the quality array to calculate an expectation score.

\renewcommand{\refname}{References}
\bibliographystyle{IEEEbib}
\bibliography{csmt2018_english}

\begin{thebibliography}{10}

\bibitem{buccoli2015unsupervised}
Michele Buccoli, Massimiliano Zanoni, Francesco Setragno, Fabio Antonacci, and Augusto Sarti,
\newblock ``An unsupervised approach to the semantic description of the sound quality of violins,''
\newblock in {\em 2015 23rd European Signal Processing Conference (EUSIPCO)}. IEEE, 2015, pp. 2004--2008.

\bibitem{park2015study}
Hyeonjun Park, Wonse Jo, Kyeongmin Choi, Hwonjae Jung, Jargalbaatar Yura, Bumjoo Lee, and Dong~Han Kim,
\newblock ``A study about sound quality for violin playing robot,''
\newblock in {\em The 10th International Conference on Future Networks and Communications {(FNC} 2015) / The 12th International Conference on Mobile Systems and Pervasive Computing (MobiSPC 2015) / Affiliated Workshops, August 17-20, 2015, Belfort, France}. 2015, vol.~56 of {\em Procedia Computer Science}, pp. 496--501, Elsevier.

\bibitem{jo2015study}
Wonse Jo, Hyeonjun Park, Bumjoo Lee, and Donghan Kim,
\newblock ``A study on improving sound quality of violin playing robot,''
\newblock in {\em 2015 6th International Conference on Automation, Robotics and Applications (ICARA)}. IEEE, 2015, pp. 185--191.

\bibitem{Suzuki2007SpectrumAA}
Hideo Hamamatsu-Shi Suzuki,
\newblock ``Spectrum analysis and tone quality evaluation of piano sounds with hard and soft touches,''
\newblock {\em Acoustical Science and Technology}, vol. 28, pp. 1--6, 2007.

\bibitem{Goebl2014PerceptionOT}
Werner Goebl, Roberto Bresin, and Ichiro Fujinaga,
\newblock ``Perception of touch quality in piano tones.,''
\newblock {\em The Journal of the Acoustical Society of America}, vol. 136 5, pp. 2839--50, 2014.

\bibitem{Palanisamy2020RethinkingCM}
Kamalesh Palanisamy, Dipika Singhania, and Angela Yao,
\newblock ``Rethinking {CNN} models for audio classification,''
\newblock {\em CoRR}, vol. abs/2007.11154, 2020.

\bibitem{Tsalera2021ComparisonOP}
Eleni Tsalera, Andreas~E. Papadakis, and Maria Samarakou,
\newblock ``Comparison of pre-trained cnns for audio classification using transfer learning,''
\newblock {\em J. Sens. Actuator Networks}, vol. 10, no. 4, pp. 72, 2021.

\bibitem{DiMaggio2022IntelligentFD}
Luigi Gianpio~Di Maggio,
\newblock ``Intelligent fault diagnosis of industrial bearings using transfer learning and cnns pre-trained for audio classification,''
\newblock {\em Sensors}, vol. 23, no. 1, pp. 211, 2023.

\bibitem{mixup}
Hongyi Zhang, Moustapha Ciss{\'{e}}, Yann~N. Dauphin, and David Lopez{-}Paz,
\newblock ``mixup: Beyond empirical risk minimization,''
\newblock in {\em 6th International Conference on Learning Representations, {ICLR} 2018, Vancouver, BC, Canada, April 30 - May 3, 2018, Conference Track Proceedings}. 2018, OpenReview.net.

\bibitem{augmix}
Dan Hendrycks, Norman Mu, Ekin~Dogus Cubuk, Barret Zoph, Justin Gilmer, and Balaji Lakshminarayanan,
\newblock ``Augmix: {A} simple data processing method to improve robustness and uncertainty,''
\newblock in {\em 8th International Conference on Learning Representations, {ICLR} 2020, Addis Ababa, Ethiopia, April 26-30, 2020}. 2020, OpenReview.net.

\bibitem{cutout}
Terrance Devries and Graham~W. Taylor,
\newblock ``Improved regularization of convolutional neural networks with cutout,''
\newblock {\em CoRR}, vol. abs/1708.04552, 2017.

\bibitem{yolov4}
Alexey Bochkovskiy, Chien{-}Yao Wang, and Hong{-}Yuan~Mark Liao,
\newblock ``Yolov4: Optimal speed and accuracy of object detection,''
\newblock {\em CoRR}, vol. abs/2004.10934, 2020.

\bibitem{Yun2019CutMixRS}
Sangdoo Yun, Dongyoon Han, Seong~Joon Oh, Sanghyuk Chun, Junsuk Choe, and Youngjoon Yoo,
\newblock ``Cutmix: Regularization strategy to train strong classifiers with localizable features,''
\newblock in {\em Proceedings of the IEEE/CVF international conference on computer vision}, 2019, pp. 6023--6032.

\bibitem{Moore1983SuggestedFF}
Brian C.~J. Moore and Brian~R. Glasberg,
\newblock ``Suggested formulae for calculating auditory-filter bandwidths and excitation patterns.,''
\newblock {\em The Journal of the Acoustical Society of America}, vol. 74 3, pp. 750--3, 1983.

\bibitem{Glasberg1990DerivationOA}
Brian~R. Glasberg and Brian C.~J. Moore,
\newblock ``Derivation of auditory filter shapes from notched-noise data,''
\newblock {\em Hearing Research}, vol. 47, pp. 103--138, 1990.

\bibitem{Peeters2011TheTT}
Geoffroy Peeters, Bruno~L. Giordano, Patrick Susini, Nicolas Misdariis, and Stephen McAdams,
\newblock ``The timbre toolbox: extracting audio descriptors from musical signals.,''
\newblock {\em The Journal of the Acoustical Society of America}, vol. 130 5, pp. 2902--16, 2011.

\bibitem{Harman1961ModernFA}
R.~W. Hiorns,
\newblock ``Modern factor analysis,''
\newblock {\em Comput. J.}, vol. 11, no. 2, pp. 219, 1968.

\bibitem{Maaten2008VisualizingDU}
Laurens Van~Der Maaten and Geoffrey~E. Hinton,
\newblock ``Visualizing data using t-sne,''
\newblock {\em Journal of Machine Learning Research}, vol. 9, pp. 2579--2605, 2008.

\bibitem{lin2017focal}
Tsung-Yi Lin, Priya Goyal, Ross~B. Girshick, Kaiming He, and Piotr Doll{\'a}r,
\newblock ``Focal loss for dense object detection,''
\newblock {\em IEEE Transactions on Pattern Analysis and Machine Intelligence}, vol. 42, pp. 318--327, 2017.

\bibitem{Simonyan2014VeryDC}
Karen Simonyan and Andrew Zisserman,
\newblock ``Very deep convolutional networks for large-scale image recognition,''
\newblock in {\em 3rd International Conference on Learning Representations, {ICLR} 2015, San Diego, CA, USA, May 7-9, 2015, Conference Track Proceedings}, Yoshua Bengio and Yann LeCun, Eds., 2015.

\bibitem{He2015DeepRL}
Kaiming He, Xiangyu Zhang, Shaoqing Ren, and Jian Sun,
\newblock ``Deep residual learning for image recognition,''
\newblock in {\em Proceedings of the IEEE conference on computer vision and pattern recognition}, 2016, pp. 770--778.

\bibitem{Huang2016DenselyCC}
Gao Huang, Zhuang Liu, Laurens Van Der~Maaten, and Kilian~Q Weinberger,
\newblock ``Densely connected convolutional networks,''
\newblock in {\em Proceedings of the IEEE conference on computer vision and pattern recognition}, 2017, pp. 4700--4708.

\bibitem{Krizhevsky2012ImageNetCW}
Alex Krizhevsky, Ilya Sutskever, and Geoffrey~E. Hinton,
\newblock ``Imagenet classification with deep convolutional neural networks,''
\newblock {\em Commun. {ACM}}, vol. 60, no. 6, pp. 84--90, 2017.

\bibitem{Zhang2017ShuffleNetAE}
Xiangyu Zhang, Xinyu Zhou, Mengxiao Lin, and Jian Sun,
\newblock ``Shufflenet: An extremely efficient convolutional neural network for mobile devices,''
\newblock in {\em Proceedings of the IEEE conference on computer vision and pattern recognition}, 2018, pp. 6848--6856.

\bibitem{Harjoseputro2020MobileNetsEC}
Andrew~G. Howard, Menglong Zhu, Bo~Chen, Dmitry Kalenichenko, Weijun Wang, Tobias Weyand, Marco Andreetto, and Hartwig Adam,
\newblock ``Mobilenets: Efficient convolutional neural networks for mobile vision applications,''
\newblock {\em CoRR}, vol. abs/1704.04861, 2017.

\bibitem{Iandola2016SqueezeNetAA}
Forrest~N. Iandola, Matthew~W. Moskewicz, Khalid Ashraf, Song Han, William~J. Dally, and Kurt Keutzer,
\newblock ``Squeezenet: Alexnet-level accuracy with 50x fewer parameters and {\textless}1mb model size,''
\newblock {\em CoRR}, vol. abs/1602.07360, 2016.

\bibitem{Szegedy2014GoingDW}
Christian Szegedy, Wei Liu, Yangqing Jia, Pierre Sermanet, Scott Reed, Dragomir Anguelov, Dumitru Erhan, Vincent Vanhoucke, and Andrew Rabinovich,
\newblock ``Going deeper with convolutions,''
\newblock in {\em Proceedings of the IEEE conference on computer vision and pattern recognition}, 2015, pp. 1--9.

\bibitem{Tan2018MnasNetPN}
Mingxing Tan, Bo~Chen, Ruoming Pang, Vijay Vasudevan, Mark Sandler, Andrew Howard, and Quoc~V Le,
\newblock ``Mnasnet: Platform-aware neural architecture search for mobile,''
\newblock in {\em Proceedings of the IEEE/CVF conference on computer vision and pattern recognition}, 2019, pp. 2820--2828.

\bibitem{Shi2015AnET}
Baoguang Shi, Xiang Bai, and Cong Yao,
\newblock ``An end-to-end trainable neural network for image-based sequence recognition and its application to scene text recognition,''
\newblock {\em {IEEE} Trans. Pattern Anal. Mach. Intell.}, vol. 39, no. 11, pp. 2298--2304, 2017.

\bibitem{Pons2019musicnnPC}
Jordi Pons and Xavier Serra,
\newblock ``musicnn: Pre-trained convolutional neural networks for music audio tagging,''
\newblock {\em CoRR}, vol. abs/1909.06654, 2019.

\end{thebibliography}

\end{document}